\documentclass[12pt]{article}
\usepackage[dvips]{graphicx}
\usepackage{pdproc}

  \makeatletter 
  \def\@cite#1{[#1]} 
  \makeatother    
\begin{document}

\renewcommand{\thefootnote}{\alph{footnote}}
\newcommand{\directprod}{\mathop{\otimes}}

\title{
 Dynamical Generation of Yukawa Couplings in
 Intersecting D-brane Models
}

\author{ NORIAKI KITAZAWA}

\address{ 
Department of Physics, Tokyo Metropolitan University, \\
1-1 Minami-Osawa, Hachioji, Tokyo 192-0397, Japan
\\ {\rm E-mail: kitazawa@phys.metro-u.ac.jp}}

\abstract{
We propose a scenario
 to obtain non-trivial Yukawa coupling matrices
 for the quark-lepton mass generation
 in supersymmetric intersecting D-brane models
 in type IIA ${\bf T}^6/{\bf Z}_2 \times {\bf Z}_2$ orientifold.
As an example,
 an explicit model is constructed in which
 all the four generations of quarks and leptons
 and two pairs of massless Higgs fields are composite. 
In this model
 non-trivial Yukawa interactions are obtained by the interplay
 between the string-level higher dimensional interactions
 among ``preons'' and the dynamics of the confinement of ``preons''.
}

\normalsize\baselineskip=15pt

\section{Introduction}

The understanding of masses and mixings of quarks and leptons
 is one of the most important problems in particle physics.
It has been expected that the string theory
 can give a solution as well as the unified description of
 the fundamental interactions including gravity.
Recent developments in the model building
 based on the intersecting D-branes
 (see Refs.\cite{BGKL,AFIRU,CSU} for essential idea)
 make the explicit and concrete discussions possible. 
Especially,
 the models with low-energy supersymmetry are interesting,
 because such models are constructed as stable solutions
 of the string theory
 (see Refs.\cite{CSU,BGO,CPS,Honecker,Larosa-Pradisi,Cvetic-Papadimitriou,Kitazawa,CLL,Kitazawa2,Honecker-Ott,Kokorelis}
  for explicit model buildings).

The structure of the Yukawa coupling matrices
 in intersecting D-brane models
 is discussed in Refs.\cite{CLS,CIM,CKL,ALS,KKMO}.
The problem is that
 we typically have
 the factorized form of the Yukawa coupling matrices,
 $g_{ij} = a_i b_j,$
 or the diagonal Yukawa coupling matrices,
 if the origin of the generation
 is the multiple intersections of D-branes.
Both structures can not give
 realistic quark or lepton masses and mixings.
The efforts in the model building,
 including many Higgs doublets\cite{ALS}, for example,
 may solve the problem,
 but the origin of the generation may have to be reconsidered.

In this article,
 we propose another scenario to have generation structure.
The generation may not be originated
 from the multiple intersections of D-branes,
 but the repetitive existence of many confining forces
 for ``preons''.
For example,
 suppose that we have ``preons''
 with some specific charge under the standard model gauge group
 which is appropriate to form one generation
 with the confining USp$(2)$ gauge interaction.
If such ``preons'' are realized in an intersecting D-brane model,
 and they belong to the fundamental representation of USp$(6)$,
 then we have composite three generations by the decomposition of
 ${\rm USp}(6) \rightarrow {\rm USp}(2)
  \times {\rm USp}(2) \times {\rm USp}(2)$
 due to the D-brane splitting.
The difference of the positions of D-branes
 for each USp$(2)$ gauge symmetry
 may affect the structure of Yukawa coupling matrices. 
In the following
 we sketch a model in which this idea is explicitly realized
 (see Ref.\cite{Kitazawa2} for complete description).
We also give an evidence that
 the resultant Yukawa coupling matrices
 in this scenario can be realistic.
\newpage

\section{The Model and Yukawa Coupling Matrices}

The configuration of the intersecting D6-branes in type IIA 
 ${\bf T}^2 \times {\bf T}^2 \times {\bf T}^2
  /{\bf Z}_2 \times {\bf Z}_2$ orientifold
 is given in Table \ref{config}.
\begin{table}[h]
\begin{center}
\caption{
Configuration of intersecting D6-branes.
All three tori are considered to be rectangular (untilted).
Three D6-branes, D6${}_4$, D6${}_5$ and D6${}_6$,
 are on top of some O6-planes.
We also have orientifold image D-brane
 for each D-brane listed in this table.
}
 \begin{tabular}{|c|c|c|}
  \hline
  D6-brane & winding number & multiplicity     \\
  \hline\hline
  D6${}_1$   & $ \quad [(1,-1), (1,1), (1,0)] \quad $ & $4$ \\
  \hline
  D6${}_2$   & $ \quad [(1,1), (1,0), (1,-1)] \quad $ & $6+2$  \\
  \hline
  D6${}_3$   & $ \quad [(1,0), (1,-1), (1,1)] \quad $  & $2+2$  \\
  \hline
  D6${}_4$   & $ \quad [(1,0), (0,1), (0,-1)] \quad $ & $12$  \\
  \hline
  D6${}_5$   & $ \quad [(0,1), (1,0), (0,-1)] \quad $ & $8$ \\
  \hline
  D6${}_6$   & $ \quad [(0,1), (0,-1), (1,0)] \quad $ & $12$ \\
  \hline
 \end{tabular}
\label{config} 
\end{center}
\end{table}

\noindent
Ramond-Ramond tadpoles are canceled out in this configuration,
 and four-dimensional ${\cal N} = 1$ supersymmetry is realized
 under the condition of $\chi_1 = \chi_2 = \chi_3 = \chi$,
 where $\chi_i = R^{(i)}_2 / R^{(i)}_1$
 and $R^{(i)}_{1,2}$ are radii for each three torus $i = 1,2,3$.
The D6${}_2$-brane system consists of
 two parallel D6-branes with multiplicities six and two
 which are separated in the second torus
 in a consistent way with the orientifold projections.
The D6${}_3$-brane system consists of
 two parallel D6-branes with multiplicity two
 which are separated in the first torus
 in a consistent way with the orientifold projections.
D6${}_1$, D6${}_2$ and D6${}_3$ branes give gauge symmetries
 of U$(2)_L=$SU$(2)_L \times$U$(1)_L$,
 U$(3)_c \times$U$(1) =$SU$(3)_c \times$U$(1)_c \times$U$(1)$
 and U$(1)_1 \times$U$(1)_2$,
 respectively.
The hypercharge is defined as
\begin{equation}
 {Y \over 2} = {1 \over 2} \left( {{Q_c} \over 3} - Q \right)
             + {1 \over 2} \left( Q_1 - Q_2 \right),
\end{equation}
 where $Q_c$, $Q$, $Q_1$ and $Q_2$ are charges of
 U$(1)_c$, U$(1)$, U$(1)_1$ and U$(1)_2$, respectively.

We break three
 USp$(12)_{{\rm D6}_4}$, USp$(8)_{{\rm D6}_5}$
 and USp$(12)_{{\rm D6}_6}$
 gauge symmetries
 to the factors of USp$(2)$ gauge symmetries
 by appropriately configuring D6-branes
 of D6${}_4$, D6${}_5$ and D6${}_6$
 (see Ref.\cite{Kitazawa2} for concrete configuration).
The resultant gauge symmetries are respectively as follows.
\begin{equation}
 \mbox{USp}(12)_{{\rm D6}_4} \longrightarrow
 \directprod_{\alpha=1}^6 \mbox{USp}(2)_{{\rm D6}_4,\alpha}
\label{gauge_D6_4}
\end{equation}
\begin{equation}
 \mbox{USp}(8)_{{\rm D6}_5} \longrightarrow
 \directprod_{a=1}^4 \mbox{USp}(2)_{{\rm D6}_5,a}
\label{gauge_D6_5}
\end{equation}
\begin{equation}
 \mbox{USp}(12)_{{\rm D6}_6} \longrightarrow
 \directprod_{\alpha=1}^6 \mbox{USp}(2)_{{\rm D6}_6,\alpha}
\label{gauge_D6_6}
\end{equation}
All of these USp$(2)$ gauge intersections
 can be naturally stronger than any other unitary gauge interactions.
If we choose $\kappa_4 M_s \sim 1$ and $\chi \sim 0.1$,
 where $\kappa_4=\sqrt{8 \pi G_N}$ and $M_s = 1 / \sqrt{\alpha'}$,
 the scales of dynamics of all USp$(2)$
 gauge interactions are of the order of $M_s$,
 and the values of the standard model gauge coupling constants
 are reasonably of the order of $0.01$ at the string scale.

\begin{figure}[htb]
\begin{center}
\includegraphics*[width=10cm]{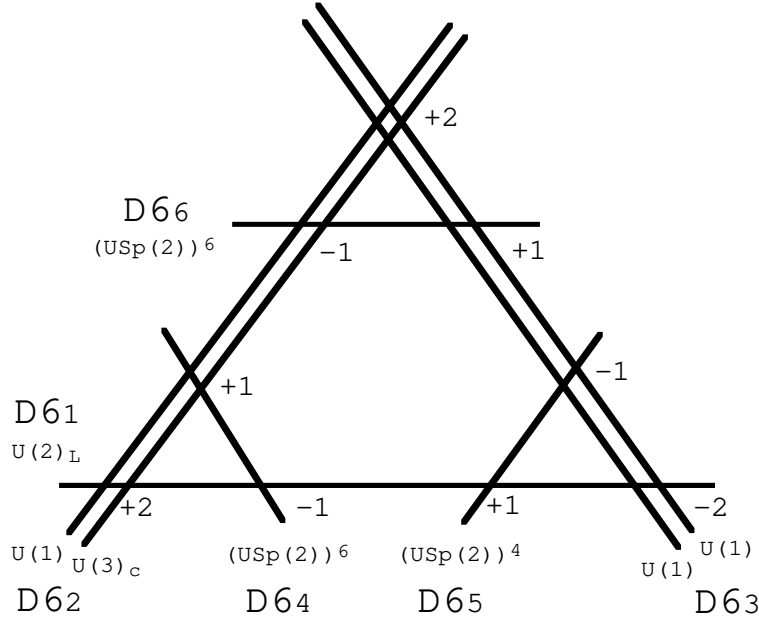}
\caption{
Schematic picture of the configuration of intersecting D6-branes.
This picture schematically shows
 the intersections of D6-branes in six-dimensional space,
 and the relative place of each D6-brane has no meaning.
The number at the intersection point
 between D6${}_a$ and D6${}_b$ branes
 denotes intersection number $I_{ab}$ with $a<b$.
}
\label{intersec}
\end{center}
\end{figure}

A schematic picture
 of the configuration of intersecting D6-branes of this model
 is given in Figure \ref{intersec}.

\noindent
``Preons'' are localized on six apices of the hexagonal area.
The sector of D6${}_1$-D6${}_2$-D6${}_4$ intersections
 gives four generation of left-handed quarks and leptons
 as two-body bound states of ``preons''.
There are six composite generations
 due to six USp$(2)$ gauge interactions of D6${}_4$-brane,
 and two anti-generations due to the twice intersection
 between D6${}_1$-brane and D6${}_2$-brane.
Two of six generations
 become massive with two anti-generations
 through the Yukawa couplings
 among two ``preons'' and one anti-generation field
 associated with the triangular aria of this sector.
The similar happens to
 the sector of D6${}_2$-D6${}_3$-D6${}_6$ intersections
 which gives four generation of right-handed quarks and leptons.
The sector of D6${}_1$-D6${}_3$-D5${}_5$ intersections
 gives two pairs of massless Higgs doublets.

The hexagonal area in Figure \ref{intersec}
 indicates the existence of
 six-point higher-dimensional interactions among ``preons''.
Since all the quarks, leptons and Higgs fields
 are two-body bound state of ``preons'',
 the higher-dimensional interactions give Yukawa interactions
 after the confinement of USp$(2)$ gauge interactions.
The value and structure of Yukawa coupling matrices
 are determined by the positions of six intersections
 of D-branes in the compact space.

In Ref.\cite{KKMO}
 the possibility to obtain non-trivial Yukawa coupling matrices
 has been shown.
The Yukawa coupling matrices of
 the heaviest two generations of up-type and down-type quarks
 in a certain condition of the sizes of three tori
 are given as follows.
\begin{equation}
g^u \sim
 \left(
  \begin{array}{cc}
  1 &
  0
  \cr
  \varepsilon_1^2 \varepsilon_3 & 
  \varepsilon_1 \varepsilon_3
  \end{array}
 \right) + {\cal O}(\varepsilon_3^2),
\qquad
g^d \sim
 \left(
  \begin{array}{cc}
  \varepsilon_1 &
  0
  \cr
  \varepsilon_1 \varepsilon_3 & 
  \varepsilon_3
  \end{array}
 \right) + {\cal O}(\varepsilon_3^2),
\end{equation}
 where $\varepsilon_i = \exp ( - A_i / 2 \pi \alpha' )$
 and $A_i$ is the $1/8$ of the area of the i-th torus.
Note that
 we can obtain Yukawa coupling of the order of unity,
 although it seems difficult
 that all six positions of intersections coincide.
These Yukawa coupling matrices
 give mass ratio and Kobayashi-Maskawa mixing angles
 for heavy two generations in a certain assumption
 to the vacuum expectation values of Higgs doublet fields.
The results are
\begin{equation}
\frac{m_{u,3}}{m_{u,4}} \sim \varepsilon_1 \varepsilon_3,
\qquad 
\frac{m_{d,3}}{m_{d,4}} \sim \frac{\varepsilon_3}{ \varepsilon_1}, \qquad 
V_{34} \sim \varepsilon_3.
\end{equation}
By taking $\varepsilon_1 \sim 0.5$ and $\varepsilon_3 \sim 0.01$,
 we obtain the values corresponding to
 $m_c/m_t \simeq 0.0038$, $m_s/m_b \simeq 0.025$,
 and $V_{cb} \simeq 0.04$.

It would be very interesting
 to explore more realistic models of the quark-lepton flavor
 in this framework.

\section{Acknowledgements}

The author thank T.~Kobayashi, N.~Maru and N.~Okada,
 for useful discussions.  

\bibliographystyle{plain}

\end{document}